\documentstyle[preprint,aps,floats,epsfig]{revtex}

\def\half{\textstyle{1\over2}}

\def\ie{{\it i.e.,}}
\newcommand{\be}{\begin{equation}}
\newcommand{\ee}{\end{equation}}
\newcommand{\bea}{\begin{eqnarray}}
\newcommand{\eea}{\end{eqnarray}}
\newcommand{\bml}{\begin{mathletters}}
\newcommand{\eml}{\end{mathletters}}

\begin{document}
\preprint{DCPT/01/41, hep-th/0104262}
\draft
\tighten



\title{Nested braneworlds and strong brane gravity}
\author{Ruth Gregory and Antonio Padilla}
\address{Centre for Particle Theory, 
         Durham University, South Road, Durham, DH1 3LE, U.K.}
\date{\today}
\setlength{\footnotesep}{0.5\footnotesep}

\maketitle
\begin{abstract}
We find the gravitational field of a `nested' domain wall living entirely
within a brane-universe, or, a {\it localised} vortex within a wall. For a 
vortex living on a critical Randall-Sundrum brane
universe, we show that the induced gravitational field
on the brane is identical to that of an $(n-1)$-dimensional vacuum
domain wall. We also describe how to set-up a nested Randall-Sundrum
scenario using a flat critical vortex living on a subcritical (adS)
brane universe.
\end{abstract}
\pacs{PACS numbers: 04.50.+h, 11.27.+d \hfill hep-th/0104262}


The old idea~\cite{RSh} that our universe might somehow be a `defect' within a
higher dimensional spacetime has recently acquired a new lease of life~--
mainly because of interest in unusual geometric resolutions of the hierarchy
problem~\cite{ADD,RS1}. 
The basic idea is that our observable 
four-dimensional universe is somehow a localised worldbrane in a
higher dimensional spacetime, and therefore, unlike the standard
Kaluza-Klein compactifications, physics is not averaged over these
extra dimensions, but strongly localised on this braneworld.
Although the original models of Arkani-Hamed et.\ al.\ \cite{ADD}
did not consistently treat gravity by including the effect of
the brane-universe itself (however see Sundrum \cite{Sun} for a codimension 2
resolution, and~\cite{CEG} for higher codimension) the feature of
generating a hierarchy between the gravitational
and other interactions via the volume factor (effective or real) of the
internal extra dimensions was nevertheless an important idea.
The Randall-Sundrum (RS) scenario on the other hand, was fully
gravitationally consistent, and saw our universe as a domain wall (or walls)
sitting at the edge of a five-dimensional anti-de Sitter spacetime.
The advantage of having a bulk adS spacetime is that although the 
extra dimension is infinite, its volume is finite, and therefore
it gives a finite contribution to the four-dimensional Planck scale.

The interesting feature of the RS scenario, and the one which makes it a viable
brane-{\it universe} model, is that is that gravity on
the domain wall is precisely Einstein gravity at low energies, \ie
\be 
R_{ab} - {1\over2} R g_{ab} = 8\pi GT_{ab} 
\ee 
this has been shown perturbatively~\cite{RS1,GT}, but so far, 
non-perturbative gravitational
solutions have been thinner on the ground.  Essentially they fall into one of
three categories: Cosmological solutions~\cite{BL,COS,BCG}, `zero-mode'
solutions which are translationally invariant orthogonal to the brane
\cite{CHR}, and gravitational wave solutions~\cite{CG}. All of these
generically contain some sort of singularity e.g.\ the cosmological solutions
tend to have initial or final singularities, and some, such as the
zero mode Schwarzschild black string solution, are unstable \cite{adsins}.
A non-perturbative solution such
as a black hole bound to the brane for example requires a five-dimensional
C-metric, which is so far unforthcoming. (See~\cite{EHM} for a lower, 3+1,
dimensional analogue.)

There is another way however, in which we could examine non-perturbative
solutions without use of the C-metric, and that is by putting an extended
source on the brane. At the linearized level, a cosmic string in the RS
braneworld (a source of codimension 3) exhibits deviations from the
pure 4-dimensional Einstein gravity \cite{Davis}, therefore one might
expect that if nonperturbative exact solutions can be found, they would provide
a way in which to explore deviations from Einstein gravity on the braneworld.

In order to explore this issue, in this paper we focus instead 
on a cosmic ``domain wall'' living entirely {\it within},
and totally localised upon, the brane universe. From the higher 
dimensional perspective, such a configuration
is that of a vortex (by which we mean  a codimension 2 object, 
rather than some solitonic solution of a field theory) 
lying totally within the wall which constitutes the
brane universe. This configuration turns out to be directly and completely
integrable, and represents a genuinely {\it fully localised}
`intersection' of the two `branes'. In the same way as one can view the RS 
scenario as a limit of a thick domain wall \cite{Gremm}, one can view this 
solution as a zero-thickness limit of a nested topological defect 
\cite{Morris}, which can occur when one has condensates of other
fields in the presence of a topological domain wall background,
poetically called a {\it domain ribbon}.

Obviously the first step is to derive such a solution and its global spacetime
structure, which we do presently, but once we have this solution, we can use
it to explore the question of the (non-perturbative) gravitational field on the
brane. Surprisingly perhaps, this turns out to look very much the same as 
if we had been ignorant of the bulk, and simply computed the gravitational 
field of {\it our} ``domain wall'' using the appropriate Einstein gravity 
in one dimension less.  
This shows that at least for these highly symmetric setups, the 
gravitational interaction on the brane universe is Einsteinian in 
nature and for the critical RS universe is proven to be Einstein
even at the non-perturbative level.

To start off, note that the gravitational field of a vortex-wall 
will have dependence on only two
spacetime coordinates, $r$ and $z$ say, with $z$ roughly
representing the direction orthogonal to the domain wall representing
our brane universe, and $r$ the direction orthogonal to the vortex
or domain ribbon within our brane universe.
We therefore expect that, schematically, the energy-momentum 
tensor of the system will have the form:
\be
T_{ab} = \sigma h_{ab} {\delta(z) \over \sqrt{g_{zz}}}
+ \mu \gamma_{ab} {\delta(z) \delta(r) \over \sqrt{g_{zz}g_{rr}}}
\label{emtens}
\ee
where $h_{ab}$ is the induced metric on the brane universe, and $\gamma_{ab}$
the induced metric on the vortex. The most general metric consistent with
these symmetries can (generalizing \cite{BCG}) in $n$-dimensions
be reduced to the form
\be
ds^2 = A^{2\over(n-2)} d{\bf x}_\kappa^2 - e^{2\nu} A^{-{(n-3)\over(n-2)}} 
(dr^2 + dz^2)
\ee
where $d{\bf x}_\kappa^2$ represents the `unit' metric on a constant curvature
spacetime ($\kappa=0$ corresponds to an $(n-2)$-dimensional
Minkowski spacetime, $\kappa = \pm 1$ to $(n-2)$-dimensional
de-Sitter and anti-de Sitter spacetimes), and the brane universe sits at $z =
0$, the vortex at $r = z = 0$. This is basically a double analytic 
continuation of the cosmological metric
in~\cite{BCG}, where it is the time translation symmetry $\partial_t$ which is
broken, rather than $\partial_r$. The key result of that paper needed here was
to show that the conformal symmetry of the $t,z$ plane meant that
the gravity equations were completely integrable in the bulk, and the 
brane universe was simply a boundary ($T(\tau), Z(\tau)$) of that
bulk (identified with another boundary of another general bulk). The 
dynamical equations of the embedding of the boundary reduced to
pseudo-cosmological equations for $Z(\tau)$. We may therefore use the results
of~\cite{BCG} (appropriately modified) to deduce that our solution must be a
section, $(R(\zeta),Z(\zeta))$ of the general bulk metric
\be\label{canbulk}
ds^2 = Z^2 d{\bf x}_\kappa^2 - h(Z) dR^2 - {dZ^2 \over h(Z)} 
\ee
where $d{\bf x}_\kappa^2$ is now a constant curvature Lorentzian spacetime, and
in general the function $h$ is
\be
h(Z) = k_n^2 Z^2 + \kappa + {c\over Z^{(n-3)}}
\ee
where $k_n^2 = -2\Lambda/(n-1)(n-2)$. If $c>0$, the metric becomes singular at
the adS horizon, $Z=0$. However, if $c<0$, the metric is analogous to a
euclidean black hole, and $R$ becomes an angular coordinate -- the spacetime
closing off before the adS horizon.

For simplicity, we will assume our
brane universe is Z$_2$ symmetric (\ie\ spacetime is reflection symmetric
around the wall) and that the integration constant, $c$, vanishes.
This gives the equations of motion for the source (\ref{emtens}) as
\bml\label{traject}\bea
Z^{\prime2}(\zeta) &=& \left ( k_n^2 - \sigma^2_n \right )
Z^2 + \kappa \label{zcos} \label{trajecta} \\
Z''(\zeta) &=& \left ( k_n^2 - \sigma^2_n \right )
Z - {\mu_n\over 2} \sigma_n Z \delta(\zeta) \label{trajectb} \\
R'(\zeta) &=& {\sigma_n Z\over (k_n^2 Z^2 + \kappa)} \label{trajectc} 
\eea\eml
where $\sigma_n = 8\pi G_n \sigma/ 2(n-2)$, and $\mu_n = 8\pi G_n\mu$.
For example, the Randall-Sundrum domain wall (in $n$-dimensions) is given
by setting $\kappa = \mu = 0$ (flat, no vortex) and $\sigma_n
= k_n$.  The bulk metric is then
\be
ds^2 = Z^2 (dt^2 - dx^2_i -k^2_n dR^2) - {dZ^2\over k_n^2Z^2} 
\ee
and we have the solution $Z=Z_0$ a constant, and $kR = \zeta/Z_0$.
Letting $Z_0=1$, and $Z = e^{-k_nz}$ gives the usual RS coordinates.
Replacing the Minkowski metric (in brackets) by an arbitrary 
$(n-1)$-dimensional metric gives the usual relation between 
Newton's constant in $n$ and $n-1$ dimensions for the RS universe:
\be\label{newton}
G_{n-1} = {(n-3)\over2} k_nG_n = {(n-3)\over2} \sigma_nG_n
\ee
a relationship confirmed by the perturbative analysis
of~\cite{RS1,GT}.

In general, the $Z$-equation (\ref{zcos}) can be integrated away from $R=0$
to give
\be\label{genzsolns}
Z = \cases{ {1\over 2\sqrt{a}} \left [
e^{\pm\sqrt{a}(\zeta-\zeta_0)} - \kappa e^{\mp \sqrt{a}(\zeta-\zeta_0)} 
\right] & $a>0$\cr
Z_0 \pm \kappa \zeta & $a=0$, $\kappa = 0,1$ \cr
{1\over \sqrt{|a|}} \cos\sqrt{|a|}(\pm\zeta - \zeta_0) & $a<0$, 
$\kappa = 1$ only \cr}
\ee 
where $a = k_n^2 - \sigma^2_n$, which is zero for a critical wall, and is
positive (negative) for a sub (super) critical wall respectively.
In the absence of the vortex, a critical wall is one with a
Minkowski induced metric, and is the original RS scenario \cite{RS1}; a
supercritical wall is one which has a de-Sitter induced metric, and
can be regarded as an inflating cosmology \cite{COS}; whereas
the subcritical wall has an adS induced metric, and has only recently
been considered from the phenomenological point of view \cite{KR}.

Since we are interested in having a domain ribbon on our brane universe, 
we require solutions with nonzero $\mu_n$, and hence a discontinuity in $Z'$.
To achieve this, we simply patch together different branches of the 
solutions (\ref{genzsolns}) for $\zeta>0$ and $\zeta<0$; 
the $R$-coordinate is given by integrating (\ref{trajectc}).

From (\ref{genzsolns}) we see that critical and supercritical walls can
only support a vortex if $\kappa=1$, \ie\ if the induced metric on the vortex
itself is a de-Sitter universe. A subcritical wall on the other hand can
support all induced geometries on the vortex. 
To investigate strong brane gravity, we focus on two specific solutions: A
domain ribbon in a Randall-Sundrum
(critical) wall; and a ``nested RS scenario'', \ie\ a flat Minkowski
domain ribbon living on a Karch-Randall (KR) subcritical adS domain wall.
If the induced gravity on the brane is indeed Einstein gravity, then we
would expect the induced metric on the braneworld to be that of a vacuum 
domain wall and the Randall-Sundrum metric respectively. In the
latter case we would also expect the $(n-2)$-dimensional 
domain ribbon to have its own localised graviton zero-mode. 
We now demonstrate this for each example in turn. 

The Randall-Sundrum universe is a critical domain wall in adS spacetime,
\ie\ it satisfies the relation $\sigma_n=k_n$. This means that a domain
ribbon on this wall {\it must} have $\kappa=1$, \ie\ a `spherical'
spatial geometry. The full spacetime is therefore the region
$Z<Z(\zeta)$:
\bml\label{critdr}\bea
Z &=& {4\over\mu_nk_n} - |\zeta| \\
k_nR &=& \mp {\half} \ln \left [ {\mu_n^2+(4-k_n\mu_n|\zeta|)^2 \over \mu_n^2 
+ 16 } \right]
\eea\eml
of the bulk (\ref{canbulk}) with $\kappa=1$.
At first sight neither the trajectory nor bulk looks like the original
RS scenario, however, the coordinate transformation
\bml\bea
k_n u &=& e^{k_nR} / \sqrt{1+k_n^2 Z^2} \\
({\tilde t},{\tilde {\bf x}}) &=& k_n u Z (\sinh t,\cosh t\ {\bf n}_{n-2})
\eea\eml
(where ${\bf n}_{n-2}$ is the unit vector in $(n-2)$ dimensions) gives
\be
ds^2 = {1\over k_n^2u^2} \left [ d{\tilde t}^2 - d{\tilde {\bf x}}^2 - du^2 
\right ]
\ee
\ie\ the familiar planar adS metric in conformal coordinates. The
trajectory (\ref{critdr}) then becomes
\be
\cases{ u = u_0 = {\mu_n\over k_n \sqrt{16+\mu_n^2}} & $\zeta<0$\cr
{\tilde{\bf x}}^2 - {\tilde t}^2 + \left ( u - {1\over 2k_n^2u_0} \right )^2
= {1\over 4 k_n^4 u_0^2} & $\zeta>0$\cr
}
\ee
the change of coordinates means that the trajectory is no longer
manifestly Z$_2$ symmetric, however, the $\zeta<0$ branch now
becomes a subset of the RS planar domain wall, specifically, the
interior of the hyperboloid
\be\label{hyprad}
{{\tilde{\bf x}}^2 - {\tilde t}^2 \over k_n^2 u_0^2}
= {16 \over k_n^2 \mu^2_n} = \left [ 2 \pi G_{n-1} \mu \right ] ^{-2}
\ee
(using (\ref{newton})). However, recall that the global spacetime structure
of a vacuum domain wall is that of two identified copies of the interior of a 
hyperboloid in Minkowski spacetime of proper radius $1/2 \pi G_{n-1} \mu$,
\cite{CGG}, therefore (\ref{hyprad}) corresponds identically with what we
would expect from $(n-1)$-dimensional Einstein gravity. 
The $\zeta>0$ branch is a hyperboloid in the bulk centered
on $u = 1/2k_n^2 u_0$ with comoving radius $1/2k_n^2 u_0$. 
As $\mu$ increases, more and more of the hyperboloid
is removed, with the spacetime `disappearing' only as
$\mu\to\infty$. Interestingly, while this is par for the 
course for a domain wall, it is completely different to
the behaviour one would expect from a vortex.

We can easily find the induced metric on the brane universe 
\be\label{indvac}
ds_{n-1}^2 = \left ( 1 - {\mu_nk_n|\zeta|\over4}\right )^2 \left [ d{\hat t}^2 -
\left({4\over\mu_nk_n}\right)^2 \cosh^2 {\mu_nk_n{\hat t}\over 4} 
d\Omega_{I\!I}^2 \right] - d\zeta^2
\ee
where ${\hat t} = 4t/\mu_nk_n$. This is {\it precisely} the metric of 
a self-gravitating domain wall of tension $\mu$ in $(n-1)$-dimensional
Einstein gravity written in Gaussian Normal coordinates \cite{VIS}. 
This can be seen from (\ref{newton}) and 
the Israel equations in $(n-1)$-dimensions for
a wall of tension $\mu$:
\be
\Delta K_{ab} = - {8\pi G_{n-1} \mu\over (n-3)} h_{ab} = 
- {k_n\mu_n\over2} h_{ab}
\ee
which is clearly the correct expression for the jump in extrinsic curvature 
at $\zeta=0$ in (\ref{indvac}).

An interesting variation on this theme 
is to consider a sub-critical instead 
of critical brane universe. A subcritical brane universe is one for which 
the tension of the brane is not sufficient to cancel the negative bulk 
cosmological constant, $|\Lambda|$, and for which the `effective' cosmological 
constant on the brane, $-2\lambda = (n-2)(n-3) (k_n^2-\sigma_n^2)$,
is still negative. From the point of view of an observer living on the
brane, the $(n-1)$-dimensional universe is adS$_{n-1}$, and a ``domain
wall'' (\ie\ ribbon) in their universe could then be an 
$(n-1)$-dimensional Randall-Sundrum scenario \ie\ an $(n-2)$-dimensional
Minkowski universe. Therefore, to set up
this nested RS scenario, we look for a planar domain ribbon (which
we expect to obey some sort of `criticality' condition analogous
to $\sigma_n=k_n$ for the original RS wall) within a subcritical brane-world,
\ie\ a $\kappa=0$ solution from (\ref{genzsolns}).
Defining $k_{n-1}^2 = k_n^2 - \sigma_n^2$, 
\be
Z = Z_0 e^{-k_{n-1}|\zeta|} \;\;\;\;\;\;
R = \pm {4\over k_n^2 \mu_n} \left( Z^{-1} - Z_0^{-1} \right)
\ee
where $\mu_n = 4k_{n-1}/\sigma_n$. Rewriting this in terms of conformal
coordinates gives
\be
R = \pm {4\over\mu_n} (u - u_0)
\ee
Each branch of this trajectory is a KR wall, which, if it were
not for the vortex at $(u_0,0)$ would reach the adS boundary at
$R = \mp 4 u_0/\mu_n$.

The induced metric on the braneworld
\be\label{indrs}
ds_{n-1} = Z_0^2 e^{-2k_{n-1}|\zeta|} [ dt^2 - dx_i^2 ] -d\zeta^2
\ee
is that of an RS universe. However, the RS universe has a strict
`criticality' relation between the tension of the brane and the
bulk cosmological constant. Here, we have
\be
k_{n-1} = 2 \pi G_n \sigma_n \mu = {4\pi G_{n-1} \mu\over (n-3)}
\ee
(using (\ref{newton}) in terms of $\sigma_n$ rather than $k_n$) 
which is precisely the RS criticality condition $\sigma_n = 
4\pi G_n \sigma /(n-2) = k_n$ adjusted for one dimension less.

Therefore, provided we identify the $n$ and $(n-1)$-dimensional 
Newton's constants using the tension of the wall, we have found that
the induced gravity of the domain ribbon agrees precisely with that
of an $(n-1)$-dimensional gravitating domain wall (either with or
without induced cosmological constant) even at the non-perturbative level.

Naturally, it would be interesting to know the full tensor structure
of gravity, both on the domain wall braneworld, as well as on the
lower-dimensional ribbon world. A full analysis of the graviton propagator
is rather involved, not only because the domain wall braneworld now
has a nontrivial trajectory through the adS bulk (which can in any case
be remedied by a judicious -- Gaussian Normal -- choice of gauge), 
but because this trajectory now contains a `kink' -- the vortex -- 
and therefore no longer respects the bulk symmetries, thus making a
simple eigenfunction expansion of the operator impossible with respect
to the usual bases.
However, it is easy to show that at least the vortex worldvolume
does have something akin to the localized zero mode of the RS braneworld. 
To do this, one can either perform the usual perturbation analysis around
the adS background with the relevant boundary (in which case very little
changes from the original analysis of Randall and Sundrum), or one can
simply replace the flat metric in the canonical bulk form
by a general Einstein metric depending only on the vortex worldvolume
coordinates:
\be
ds^2 = Z^2 g_{\mu\nu} (x) dx^\mu dx^\nu 
- k_n^2 Z^2 dR^2 - {dZ^2\over k_n^2 Z^2 }
\ee
Interestingly however, integrating out 
the Einstein action for this zero mode now gives
\be
G_{n-2} = {(n-3)(n-4)\over 4} {k_n^2\over \sigma_n^2} G_n k_{n-1}\sigma_n
\ee
which has a discrepancy of $k_n^2 / \sigma_n^2$ over what might have 
been expected from (\ref{newton}). 
Clearly a more detailed analysis is required.

Finally, one can envisage more complicated braneworld
domain ribbon configurations than the ones described here. The general
scenario of Randall and Sundrum has of course been extended to obtain
a plethora of brane universe models many of which contain negative tension
walls. The general approach described here
can be modified to allow for negative tension walls, as
well as patching between spacetimes with different cosmological
constants, although the nontrivial trajectories induced by the
presence of the vortex will, in general, mean that either a 
``mirror'' vortex must be introduced on the negative tension brane, or
the positive tension brane must match to the negative tension brane
across the vortex.  
One suspects however, that the more baroque the model, the less stable
or useful it is likely to be.

\section*{Acknowledgements}

We would like to thank Filipe Bonjour, Ian Davies, Roberto Emparan, James
Gregory, Karen Lovis, Davina Page and Simon Ross for useful
discussions. R.G.\ was supported by the Royal Society and A.P.\
by PPARC.

\end{document}